\documentclass[12pt,preprint]{aastex}

\let\ni=\noindent
\let\new=\newcommand
\new{\scrf}{{\cal F}}
\new{\scrr}{{\cal R}}
\new{\half}{{\frac{1}{2}}}
\new{\cm}{{\rm\, cm}}
\new{\m}{{\rm\, m}}
\new{\pc}{{\rm\, pc}}
\new{\kpc}{{\rm\, kpc}}
\new{\km}{{\rm\, km}}
\new{\seco}{{\rm\, s}}
\new{\yr}{{\rm\, yr}}
\new{\Myr}{{\rm\, Myr}}
\new{\Hz}{{\rm\, Hz}}
\new{\kHz}{{\rm\, kHz}}
\new{\MHz}{{\rm\, MHz}}
\new{\GHz}{{\rm\, GHz}}
\new{\kms}{{\rm\, km\,s^{-1}}}
\new{\erg}{{\rm\, erg}}
\new{\G}{{\rm\, G}}
\new{\K}{{\rm\, K}}
\new{\msun}{{M_{\odot}}}
\new{\rsun}{{R_{\odot}}}
\new{\lsun}{{L_{\odot}}}
\new{\temperature}{{\left(\frac{T}{10^4\K}\right)}}
\new{\invtemperature}{{\left(\frac{10^4\K}{T}\right)}}
\new{\outerscale}{{\left(\frac{\ell}{1\pc}\right)}}
\new{\invouterscale}{{\left(\frac{1 \pc}{\ell}\right)}}
\new{\frequency}{{\left(\frac{\nu}{1\GHz}\right)}}
\new{\invfrequency}{{\left(\frac{1\GHz}{\nu}\right)}}
\new{\thetaobs}{{\left(\frac{\theta_{\rm obs}}{1\arcsec}\right)}}
\new{\dist}{{\left(\frac{d}{3\times 10^9\cm}\right)}}
\new{\as}{{\rm\, arcsec}}
\new{\F}{{\, \cal F}}

\shorttitle{Radio--Wave Scattering in the Galactic Center}
\shortauthors{P. Goldreich \& S. Sridhar}

\begin{document}

\title{Folded Fields as the Source of Extreme Radio--Wave Scattering
in the Galactic Center}


\author{Peter Goldreich\altaffilmark{1} and S. Sridhar\altaffilmark{2}}
\altaffiltext{1}{School of Natural Sciences, Institute for Advanced Study, 
    Einstein Drive, Princeton, NJ~08540; Email:~{\tt pmg@ias.edu}}
\altaffiltext{2}{Raman Research Institute, C.~V.~Raman Avenue,
Sadashivanagar, Bangalore 560080, INDIA; Email:~{\tt ssridhar@rri.res.in}}

\newpage
\begin{abstract}
A strong case has been made that radio waves from sources within about
half a degree of the Galactic Center undergo extreme diffractive
scattering.  However, problems arise when standard (``Kolmogorov'')
models of electron density fluctuations are employed to interpret the
observations of scattering in conjunction with those of free--free
radio emission.  Specifically, the outer scale of a Kolmogorov
spectrum of electron density fluctuations is constrained to be so
small that it is difficult to identify an appropriate astronomical
setting. Moreover, an unacceptably high turbulent heating rate results
if the outer scale of the velocity field coincides with that of the
density fluctuations.  We propose an alternative model based on folded
magnetic field structures that have been reported in numerical
simulations of small--scale dynamos. Nearly isothermal density
variations across thin current sheets suffice to account for the
scattering. There is no problem of excess turbulent heating because
the outer scale for the velocity fluctuations is much larger than the
widths of the current sheets. We speculate that interstellar magnetic
fields could possess geometries that reflect their origins: fields
maintained by the galactic dynamo could have large correlation
lengths, whereas those stirred by local energetic events might exhibit
folded structures.
\end{abstract}

\section{Introduction}

It appears that radio waves from sources within about half a degree of
the Galactic Center (GC) undergo extreme diffractive
scattering. Observations of SgrA* \citep{dav76,rogers94} and of maser
spots in several OH/IR stars \citep{van92,fra94} have established that
the angular broadening is $\theta_{\rm obs}\approx 1\arcsec$ at
$\nu\simeq 1\GHz$. If the scattering region is located more than a few
$\kpc$ from the GC, this would correspond to enhanced, but not
unusually large, levels of scattering; NGC 6334B is the record
holder with a scatter broadened disk of size $\approx 7\arcsec$ at
$\nu\simeq 1\GHz$ \citep{mor90}. However, if the scattering region is
located close to the GC (say, within $1\deg$, corresponding to a
distance $R\approx 150\pc$ from the GC), then the GC region would be a
site of extreme scattering. \citet{van92} argue that, if the
scattering region is also responsible for the free-free absorption
toward the GC, then upper limits on the optical depth
\citep{ped89,ana91} constrain $R$ to be in the range $0.85-3\kpc$. In
more recent work, \cite{laz98} make the case that $R\approx
130\pc$. Since the Sun's distance from the GC is $\approx 8.5\kpc$,
this would imply that the angular scattering is larger than the
observed angular broadening by a factor $8500/130\approx 65$:
\begin{equation}
\theta_{\rm scat} \approx 65\,\scrr^{-1}\,\theta_{\rm obs} \sim 3\times
10^{-4}\,\scrr^{-1}\nu_9^{-2}\,, 
\label{eq:thetascat}
\end{equation}
\ni where we have set $R = 130\, \scrr\pc$, and $\nu_9\equiv
\nu/(\GHz)$.  \citet{spe92} estimate the gas pressure in the Galactic
bulge to be $p\equiv nT\sim 5\times 10^6\K \cm^{-3}$. Studies based on
X-ray emission from the GC \citep{mun04} suggest that, even on the
smaller spatial scales of interest to us, the gas pressure may not be
very different. A pressure of $5\times 10^6\K \cm^{-3}$ is about
$10^{2.5}$ times higher than in the local interstellar medium, hence
it may not seem unreasonable to expect the GC to be a region of
extreme scattering for radio waves. However, as LC98 demonstrate,
problems arise when standard models of electron density fluctuations
are employed to interpret the observations of scattering in
conjunction with those of free--free radio emission.\footnote{Similar, but
less severe, problems arise for NGC 6334B which lies behind a galactic HII
region \citep{mor90}.}

\section{Problems With The Standard Interpretation}

LC98 estimated the brightness temperature of free-free emission toward five
highly scattered OH maser sources near the GC, based on the $10\GHz$
survey of \citet{han87}. From Table~3 of LC98, we derive a mean
brightness temperature, $T_b \approx 0.25\K\,$ at $\nu=10\GHz\,$. Since
the region is optically thin, the optical depth $\tau_{\rm ff}\simeq
(T_b/T) \approx 2.5\times 10^{-5}T_4^{-1}$, where $T_4\equiv T/(10^4\K)$
is the gas temperature. Expressed in terms of the emission measure ($EM$),
$\tau_{\rm ff}\approx 5\times 10^{-7} \nu_9^{-2} T_4^{-3/2} EM_1$, where
$EM_1\equiv EM/(\pc\cm^{-6})$, and we have set the averaged Gaunt
factor ${\overline g_{\rm ff}}\approx 10$. Hence,
\begin{equation} 
EM \sim 5\times 10^3\,T_4^{1/2}\,\pc\cm^{-6}. 
\label{eq:EM} 
\end{equation} 
\ni The emission measure contributed by a medium with pressure $p =
5\times 10^6\,\scrf\K \cm^{-3}$, is $EM \sim fn^2R \approx 3.3\times
10^7\,f\,\scrf^2\,T_4^{-2}\,\scrr\,\pc\cm^{-6}$, where $f$ is the
volume filling factor. Provided all the gas that contributes to the EM
is at temperature $T_4$, $f\sim 1.5\times 10^{-4}T_4^{5/2}\,
\scrf^{-2}\,\scrr^{-1}$ and the mass of this gas is $M\sim 1.5\times
10^4\, T_4^{3/2}\,\scrf^{-1}\,\scrr^2\;\msun$. Obviously the
constraint $f\lesssim 1$ implies $T_4\lesssim
30\,\scrf^{4/5}\,\scrr^{2/5}$. 

In the standard interpretation of radio wave scattering in the
interstellar medium, the electron density fluctuations are assumed to
follow a Kolmogorov spectrum, $(\Delta n_\ell/n)\approx
(\ell/L)^{1/3}$ for $\ell_{\rm min}\leq \ell\leq L$. For SgrA*,
$\theta_{\rm scat}\propto \nu^{-2}$ for $\nu_9\lesssim 30$. The
$\nu^{-2}$ scaling implies that the scales contributing dominantly to
strong, diffractive scattering are smaller than the inner scale:
$\ell_{\rm min} > \lambda/\theta_{\rm scat}\, $, where $\lambda =
c/\nu$ is the wavelength of the radio waves.\footnote{If scattering
was dominated by scales between $\ell_{\rm min}$ and $L$, we would
have $\theta_{\rm scat}\propto\nu^{-2.2}\,$.}  Because
$\lambda/\theta_{\rm scat} \propto \nu$, we obtain a lower limit on
$\ell_{\rm min}$ which is higher at higher frequency. Using
equation~(\ref{eq:thetascat}), with $\nu_9=30$, gives $\ell_{\rm min}
\geq 3\times 10^6\,\scrr\cm$. The scattering angle, $\theta_{\rm scat}
\sim (\nu_p/\nu)^2 (fR/L)^{1/2} (L/\ell_{\rm min})^{1/6}$, where
$\nu_p \approx 10^4(n/\cm^{-3})^{1/2}\Hz$ is the plasma frequency. In
this expression for $\theta_{\rm scat}$, we can write $fR\approx
EM/n^2$. Using equations~(\ref{eq:thetascat}) and (\ref{eq:EM}) for
$\theta_{\rm scat}$ and $EM$, the condition $\ell_{\rm min} \geq
3\times 10^6\,\scrr\cm$ implies that,
\begin{equation}
L\lesssim 3\times 10^{-8}T_4^{3/4}\,\scrr^{5/2}\pc.
\label{eq:lmax}
\end{equation}
\ni The upper limit to $L$ is remarkably small and independent of the
pressure.\footnote{The problem of a small outer scale for density
fluctuations arises more generally. \citet{ana88} estimated that
enhanced scattering in the inner galaxy required fractional density
fluctuations $\sim 10$ on the outer scale. For a Kolmogorov spectrum,
the same enhancement may be achieved with order unity fractional
density fluctuations, with an outer scale that is $10^3$ times
smaller. In this case, the outer scale of the velocity fluctuations
can be much larger than the outer scale of the density fluctuations.}
Moreover, if only a fraction $F$ of the plasma that contributes to the
free-free emission is responsible for the scattering, $L$ would be
smaller by a factor $F^{3/2}$. This, in essence, is LC98's
argument. Henceforth we set $\scrf = \scrr =1$.

The choice of a small outer scale, constrained by
equation~(\ref{eq:lmax}), introduces two problems.  As was realized by
LC98, it implies an unacceptably high heating rate if $L$ is also the
outer scale of the velocity fluctuations. Dissipation by Kolmogorov
turbulence would heat the gas at rate, $t_{\rm turb}^{-1}\sim c_s/L >
10^{-5}T_4^{-1/4}\seco^{-1}$, whereas radiative cooling would be much
slower, $t_{\rm cool}^{-1}\sim 3\times 10^{-7}T_4^{-2} \Lambda_{-21}
\seco^{-1}$ (here $c_s\sim 10T_4^{1/2}\kms$ is the sound speed, and
$\Lambda_{-21} \equiv \Lambda/(10^{-21} \erg \seco^{-1}\cm^3)$;
$\Lambda_{-21}\approx 1$ when $T_4\approx 10$). It is also difficult
to identify an appropriate astronomical setting in which nonlinear
density fluctuations on a scale as small as $L$ might arise. We have
considered radiative shocks and interfaces between neutral and ionized
gas formed by ionizing radiation or hot gas incident upon the surfaces
of a molecular clouds. Even at the high pressure in the GC region,
each case yields a length scale that is at least a few orders of
magnitude larger than $L$.

\section{Folded Fields As A Solution}

Folded magnetic field structures appear in numerical simulations of
small--scale, turbulent dynamos~\citep{sch04}. Of relevance here are
the electrical conductivity, $\sigma\approx 10^{13}T_4^{3/2}
\seco^{-1}$, and the kinematic viscosity, $\nu_{\rm vis}\approx
2\times 10^{15}T_4^{7/2} \cm^2\seco^{-1}$, for plasma with gas
pressure, $p\approx 5\times 10^6\K \cm^{-3}$.\footnote{These scalar
transport coefficients only apply in directions parallel to the local
magnetic field. } The magnetic Prandtl number, ${\rm Pr}_m\equiv 4\pi
c^{-2}\sigma\nu_{\rm vis}\approx 3\times 10^8 T_4^5$, is large so
dissipation of velocity fields occurs on much larger scales than
dissipation of magnetic fields. \citet{sch04} propose a model for the
organization of magnetic fields that appears attractive; we describe
this briefly below, before considering its implications for density
fluctuations.

Let an incompressible fluid, permeated by a weak magnetic field, be
stirred on an outer scale $L$, with random velocity $v_L$. Within a
few stirring times, $\tau_L\sim (L/v_L)$, the kinetic energy cascades
turbulently, and creates velocity fluctuations on smaller spatial
scales through nonlinear hydrodynamic interactions. The rms velocity
across a scale $\ell$ may be expected to follow a Kolmogorov spectrum,
$v_\ell\sim v_L(\ell/L)^{1/3}$, for $\ell_{\rm vis}\leq\ell\leq L$,
where the inner scale is $\ell_{\rm vis}\sim L(\nu_{\rm
vis}/Lv_L)^{3/4} \ll L$. Eddies of scale $\ell$ turn over on times,
$\tau_\ell\sim (\ell/v_\ell)\sim \tau_L(\ell/L)^{2/3}$.  Hence the
early evolution of the weak magnetic field will be dominated by the
stretching action of the smallest eddies, of size $\sim\ell_{\rm
vis}$, because their turn--over time is the shortest. These eddies
cease to be effective at stretching the field lines when the magnetic
energy density becomes comparable to their kinetic energy density,
i.e. when $B^2\sim m_pnv_{\rm vis}^2$. Then larger and more energetic
eddies take over and the magnetic energy density continues to grow
until it achieves approximate equipartition with the kinetic energy of
the largest eddies: $B^2\sim m_pnv_L^2$. It is the geometry of the
magnetic field that is of particular interest to us. According to
\citet{sch04}, it has a folded structure, with parallel correlation
length $\sim L$: after a distance $L$, a typical field line reverses
direction sharply, and folds back on itself. Sheets of
direction--reversing folded fields of thickness $d$ are separated
by current sheets of similar thickness,\footnote{Folded fields
in cartoon form, and as produced in a simulation, are shown in Figs.~10 \&
15 of \citet{sch04}.}
\begin{equation}
d\sim \left({c^2\over 4\pi\sigma}\tau_L\right)^{1/2}\,
\sim 10^{10}T_4^{-3/4}\tau_6^{1/2}\cm\,,
\label{eq:d}
\end{equation}
\ni where $\tau_6\equiv\tau_L/(1\Myr)\,$. For an isothermal gas at
fixed total pressure, the fractional density perturbation across a
current sheet of thickness $d$ is,
\begin{equation}
{\Delta n\over n}\sim {B^2\over 8\pi nkT}\equiv \beta^{-1}.
\label{eq:beta}
\end{equation}

Next we estimate scattering of radio waves by folded fields.  Consider
the idealized case of an ensemble of plane--parallel current sheets,
each of radius $L$ and thickness $d\ll L$. Let the sheets fill space
statistically homogeneously with filling factor, $f$, and be oriented
randomly.\footnote{A modest non--random orientation could account
for the anisotropic images of scatter broadened sources near the GC.}
The rms phase difference across a transverse scale $d$, accrued along
a path length $R\gg L$, can be estimated by imagining a ray going
through $(R/L)$ independently oriented plates, of which only a
fraction $(d/L)$ is oriented favorably enough to each contribute to
the phase a path length $\sim L$:
\begin{equation}
\Delta\Phi \sim \left(\frac{\nu_p}{\nu}\right)^2\,\frac{\Delta n}{n}\,
\frac{L}{\lambda}\,\left(f\,\frac{R}{L}\,\frac{d}{L}\right)^{1/2}
=\left(\frac{\nu_p}{\nu}\right)^2\,\frac{\Delta n}{n}\,
\left(f\,\frac{R\,d}{\lambda^2}\right)^{1/2}\,.
\label{eq:Phi}
\end{equation}
\ni For a more realistic case, we take the sheets to have radii of
curvature $r_c$, perhaps caused by Alfv\'en waves which can propagate
along folded fields. To estimate the rms phase difference for $r_c > L$,
there are two limits to consider. For $L < \sqrt{dr_c}\,$, the situation
is similar to that for plane--parallel sheets, and $\Delta\Phi$ is given
by equation~(\ref{eq:Phi}). For $ \sqrt{dr_c} < L < r_c$, a ray goes
through $(R/L)$ independently oriented plates, of which a larger fraction
$(L/r_c)$ is oriented favorably enough for each plate to contribute path
length $\sim \sqrt{dr_c}\,$. Therefore, the expression for $\Delta\Phi$
remains unchanged, so long as $r_c > L$. We can also arrive at the above
conclusions more formally by estimating the phase structure function, due
to isotropically oriented sheets (see Appendix). The angular scattering,
$\theta_{\rm scat}\sim (\lambda/d)\Delta\Phi \sim
(\nu_p/\nu)^2(fR/d)^{1/2}\beta^{-1}$, depends on $L$ through $d$ but is
otherwise independent of $L$. We write $fR\approx (EM/n^2)$, and use
equation~(\ref{eq:d}) for $d$, to obtain a general expression for the
angular scattering:
\begin{equation}
\theta_{\rm scat}\sim \left(\frac{\nu_p}{\nu}\right)^2
\left(\frac{EM}{n^2d}\right)^{1/2}\beta^{-1}\,\sim 2\times 10^{-6}\,
EM_1^{1/2}T_4^{3/8}\tau_6^{-1/4}\nu_9^{-2}\beta^{-1}\,.
\label{eq:thscfold}
\end{equation}
\ni This is our main result. It is independent of the geometrical
distribution of the scattering material, but this deserves further
attention.

The scattering material must cover most of the area as seen from outside
$R$. Let us suppose that folded fields fill an approximately spherical
shell of radius $R$ and thickness
\begin{equation} 
\Delta R \sim fR\sim 2\times 10^{-2}T_4^{5/2}\,\scrf^{-2}\,\pc\,.
\label{eq:deltaR} 
\end{equation} 
Assuming that turbulent
stirring occurs at sonic speeds, $v_L\sim
10\,T_4^{1/2}\kms$, we find 
\begin{equation}
\tau_6\sim 2\times 10^{-3}T_4^2\left(L\over \Delta R\right)\,.
\label{eq:tau6num}
\end{equation}
\ni Substituting this value of $\tau_6$ and equation~(\ref{eq:EM}) for $EM$
in equation~(\ref{eq:thscfold}), we obtain,
\begin{equation}
\theta_{\rm scat}\sim 5\times 10^{-4}\left(T_4^{1/8}\over
\nu_9^2\beta\right)\left(\Delta R\over L\right)^{1/4}\,,
\label{eq:thscdnum}
\end{equation}
\ni which compares well with the value of $\theta_{\rm scat}$, derived from
observations, given by equation~(\ref{eq:thetascat}). The thickness of the
current sheets can be estimated by substituting
equation~(\ref{eq:tau6num}) for $\tau_6$ in equation~(\ref{eq:d}):
\begin{equation}
d\sim 6\times 10^8 T_4^{1/4}\left(L\over \Delta R\right)^{1/2}\cm\,. 
\label{eq:dnum}
\end{equation}
\ni The scattering is strong because 
\begin{equation}
\Delta \Phi\sim {d\over\lambda}\theta_{\rm scat}\sim 10^4
\left(T_4^{3/8}\over \nu_9\beta\right)\left(L\over \Delta
R\right)^{1/4}\,.
\label{eq:DelPhinum}
\end{equation}
\ni One motivation for considering scattering from a shell of warm ionized
gas is the presence of a lobe of emission surrounding the GC, discovered
by \citet{sof84} in a $10\GHz$ survey. Comparison with the $5\GHz$ survey
of \citet{alt78} enabled them to decompose the emission into thermal and
nonthermal components, and establish that the thermal component arises
mostly from a shell-like feature \citep{sof85}. This is one of the larger
of the many sources in the GC region \citep{lar00}, with an angular size
exceeding a degree. Even so, its radius $\approx 80\pc$ appears to be only
a little more than half the value of $\approx 130\pc$ we assumed, based on
LC98's location of the scattering region. This implies that $\Delta R$
might be only a little larger than half the value given in
equation~(\ref{eq:deltaR}). However, our estimates of $\theta_{\rm scat}$
and $\Delta\Phi$ given above are hardly affected, because of their weak
dependence on $\Delta R$.

\section{Summary} 

Turbulence giving rise to folded fields may account for extreme scattering
in the GC region because large fractional density fluctuations occur
across the thickness of the current sheets, $d$, which is much smaller
than the outer scale, $L$. Although the current sheets are thin,
over-heating is not a problem. The ohmic dissipation rate per volume due
to current sheets is $\sim (cB/d)^2\sigma^{-1}\sim B^2/\tau_L$. If, as we
have argued, the magnetic energy achieves near-equipartition with the
kinetic energy, then $B^2\sim m_pnv_L^2$, and the ohmic dissipation rate
is no larger than the turn--over rate of the largest eddies: $t_{\rm
ohm}^{-1}\sim \tau_L^{-1}$, which is much smaller than $t_{\rm
cool}^{-1}$. The large cooling rate also ensures that the plasma behaves
nearly isothermally, an assumption used in our estimate of density
fluctuations across the thickness of the current sheets, given in
equation~(\ref{eq:beta}).

In our discussion of radio--wave scattering we assumed that the
current sheets are oriented isotropically. It then proves convenient
to formulate radio--wave scattering in terms of angle--averaged
quantities. The effective, isotropic power spectrum of electron
density fluctuations turns out to be shallow, being proportional to
$k^{-2}$ (see equation~\ref{eq:P}). Such a spectrum gives rise to
large density fluctuations on spatial scales $\sim d$, but contributes
little to density fluctuations on large spatial scales. This
difference provides an observational test of the folded field
hypothesis. Where diffractive and refractive scintillations have been
detected in the same source, their relative magnitudes suggest that a
Kolmogorov spectrum of density fluctuations extends up to at least the
refractive scales. However, refractive scintillations are generally
not detected in strongly scattered sources. This does not conflict
with the Kolmogorov model which predicts them to be both weak and
slow, but it is also compatible with the folded field
hypothesis. Higher frequency observations could provide a decisive
test. For a Kolmogorov spectrum, the fractional flux modulation, $m$,
is of order the cube--root of the ratio of the diffractive scale to
the Fresnel scale. With parameters appropriate to SgrA*, we find
$m\sim 10^{-2}\scrr^{1/6}\nu_9^{1/2}$.

Magnetic fields might possess different geometries in different
regions of the interstellar medium, depending on their origins. Some
regions could possess strong, mean magnetic fields, correlated over
large distances, generated, perhaps, by large--scale dynamos. These
could be the sites of anisotropic Kolmogorov turbulence \citep{gol95},
responsible for the general level of diffractive and refractive
interstellar scintillation. There could be other sites in the
interstellar medium, permeated with folded fields, generated by
small--scale dynamos, contributing to extreme diffractive scattering,
but little to refractive scattering: the Galactic Center could be one
such region.

We would like to inject a cautionary concluding note. Our model of
scattering applies results on the growth of magnetic fields in
small--scale dynamos from \citet{sch04}. These are based on MHD with
scalar diffusivities. It is an open question as to whether they can be
applied to the low density, magnetized plasma in the GC region. We expect
that our estimate for magnetic diffusivity, which sets the thickness of
the current sheets, is valid. But the reduction of the kinematic viscosity
in directions perpendicular to the magnetic field is a concern. It is
possible that folded fields would unwind so rapidly that they could not be
maintained.

\appendix

\section{Diffractive scattering by an ensemble of current sheets}

Here we offer a physical derivation of equation~(\ref{eq:Phi}), by
estimating the phase structure function due to randomly distributed and
isotropically oriented current sheets. A useful intermediate step is to
calculate $C(r)$, the effective isotropic density--density correlation
function on separation $r$. Its Fourier transform is $P(k)$, the effective
isotropic power spectrum of density fluctuations. Consider an emsemble of
flat current sheets, each of thickness $d$ and radius $L\gg d$, all
oriented in the same direction with unit normal $\hat{{\bf n}}$. The
density--density correlation function is significantly non--zero only when
the separation, ${\bf r}$, is such that $|{\bf r}\cdot\hat{{\bf n}}| <
d\,$, and $ r_{\perp} = |{\bf r} - \hat{{\bf n}}({\bf r}\cdot\hat{{\bf
n}})| < L\,$. When averaged over all directions of $\hat{{\bf n}}$, we
will obtain $C(r)$. Equivalently, we may keep $\hat{{\bf n}}$ fixed, and
average over all directions of ${\bf r}$. When $d < r < L$, the solid
angle of the region of intersection between the plate--like region (of
radius $L$ and thickness $d$), and a concentric sphere of radius $r$, is
$\sim (dr/r^2)\sim (d/r)$. Hence, 
\begin{equation}
C(r) \;\sim\; (\Delta n)^2\,\frac{d}{r}\,;\qquad\mbox{for $d < r < L$}.
\end{equation}
\noindent The corresponding effective isotropic power spectrum is
``shallow'': 
\begin{equation}
P(k) \;\sim\; (\Delta n)^2\frac{d}{k^2}\,;\qquad\mbox{for $L^{-1} < k <
d^{-1}$}.
\label{eq:P}
\end{equation}
\noindent $P(k)\sim (\Delta n)^2L^2d$ when $kL < 1$ and $P(k)\sim 0$ for
$kd > 1$. Let us consider the case when the current sheets are curved,
with radius of curvature, $r_c > L$. The solid angle of the region of
intersection between the curved current sheet--like region and a
concentric sphere of radius $r$ (in the space of separations) is, $\sim
(dr_{\perp}/r^2)$. We estimate, $r_{\perp} \simeq r\,(1 -
r^2/8r_c^2)\simeq r$, because $r/r_c < L/r_c < 1\,$. Therefore the solid
angle of intersection is $(d/r)$, and the expressions for $C(r)$ and
$P(k)$ are as given above. As may be seen, the dominant contribution to
density fluctuations comes from scales close to $d$.   

In the thin--screen model of radio--wave scattering, the entire effect of
the interstellar medium is specified by the (gaussian) random phase
pattern imprinted on a wave front, as it passes through a ``phase
screen'', placed between the source and the observer. The statistical
properties of the screen are completely described by the phase structure
function, $D({\bf s})$, which is defined as the mean square phase
difference across transverse separation ${\bf s}$ on the screen. When the
power spectrum is isotropic, $D$ depends only on $s=|{\bf s}|$:
\begin{equation}
D(s)\;=\; \frac{R}{\pi}\lambda^2r_e^2\,\int_0^\infty
dk_\perp\,k_\perp\left[1 - J_0(k_\perp s)\right]\,P(k=k_\perp)\,,
\end{equation}
\ni where $r_e = e^2/m_ec^2$ is the classical electron radius. For the
shallow spectrum of equation~(\ref{eq:P}), it is straightforward to make
the estimate,
\begin{equation}
D(s) \;\sim\; \cases{(\Delta\Phi)^2(s/d)^2, &if $s<d$;\cr
(\Delta\Phi)^2, &if $s>d$,\cr}
\end{equation}
\ni where $\Delta\Phi$ is given by equation~(\ref{eq:Phi}). If $\Delta\Phi
< 1\,$, the scattering is weak, whereas a range of scales (smaller than,
and of order $d$) can contribute to strong scattering, when $\Delta\Phi >
1\,$.\footnote{We note that $D(s)\propto s^2$ yields a Gaussian image, as
observed. A Kolmogorov spectrum of density fluctuations also yields a
Gaussian image, if the diffraction scale is smaller than the inner scale.}

\end{document}